\documentclass{IEEEtran}
\usepackage[square,sort,comma,numbers]{natbib}
\usepackage{color,array}
\usepackage{graphicx}
\usepackage{academicons}
\usepackage{amsmath}
\usepackage{xcolor}
\usepackage{longtable}
\usepackage{caption}
\usepackage{amsmath,amsfonts}
\usepackage{array}
\usepackage{url}
\usepackage{verbatim}
\usepackage{graphicx}
\usepackage{hyperref}
\usepackage{academicons}
\usepackage{amsmath}
\usepackage{xcolor}
\usepackage{longtable}
\usepackage{caption}
\usepackage{academicons}
\usepackage[skip=3pt]{caption}
\usepackage{multirow}
\usepackage{adjustbox}
\usepackage{rotating}
\usepackage{pgfplots}
\usepackage{subcaption}
\usepackage{neuralnetwork}
\usepackage{tabularx}
\usepackage{graphicx,subcaption}
\usepackage{algorithm}
\usepackage{algpseudocode}

\usepackage{tabularx}
\usepackage{bm}
\usepackage{braket}
\usepackage{lettrine}
\begin{document}

\title{A Quantum-Classical Hybrid Framework for Multivariate Time-Series Forecasting: Complexity-Fidelity Trade-offs and Limitations}
\author{Sanjay Chakraborty}
\author{Sanjay Chakraborty*,\thanks{*Sanjay Chakraborty is working in the Department of Computer and Information Science (IDA), REAL, AIICS,  at Linköping University, Sweden and Department of Computer Science \& Engineering, Techno International New Town, Kolkata, India (Email: sanjay.chakraborty@liu.se).}
\and Fredrik Heintz\thanks{Fredrik Heintz is working in the Department of Computer and Information Science (IDA), REAL, AIICS, at Linköping University, Sweden, Email: fredrik.heint@liu.se}}

\maketitle

\begin{abstract}
This paper presents a unified quantum-classical hybrid framework for multi-horizon time-series forecasting, introducing two model variants — a Quantum Reservoir Forecaster (QRC-F) and a Variational Quantum Forecaster (VQF-F), evaluated in multivariate settings. The framework further provides a systematic complexity--fidelity trade-off analysis with important limitations under near-term NISQ hardware constraints. The proposed framework first applies uniform binary quantization to map continuous time-series signals into discrete binary representations, which are subsequently encoded into quantum states via angle encoding using parameterized $R_Y$ rotation gates, with cross-channel entanglement layers explicitly capturing inter-variable dependencies in the multivariate case. QRC-F employs a fixed random unitary reservoir for stable, gradient-free temporal feature extraction, while VQF-F deploys a variational quantum circuit of depth $D$, trained through the parameter-shift rule, to extract rich temporal and inter-channel features as Pauli observable expectation values, both replacing the quadratic self-attention mechanism with linear matrix operations that significantly reduce computational complexity and model parameters. A shared MIMO-based multi-horizon readout head maps the extracted quantum features simultaneously to all prediction horizons $H$, avoiding the error accumulation inherent in recursive forecasting strategies. The predicted quantum states are subsequently decoded back into classical signals through an inverse angle mapping and de-quantization procedure, faithfully reconstructing the forecasted time-series values at the original scale. Experimental evaluations across benchmark datasets including ETTh1, ETTh2, ETTm1, ETTm2, Weather, electricity, and exchange-rate demonstrate that VQF-F offers superior training stability and parameter efficiency for multi-horizon tasks, while QRC-F provides superior training stability and circuit fidelity under hardware noise, collectively establishing a principled and practical quantum-native forecasting paradigm suitable for deployment on near-term Noisy intermediate-scale quantum (NISQ) devices.
\end{abstract}


\maketitle

\section{Introduction}
\lettrine{A}{lthough} current time-series forecasting methods have significantly advanced the state-of-the-art (SOTA), particularly for multivariate time-series forecasting, they continue to face fundamental challenges in effectively capturing multivariate long-range temporal dependencies and extracting meaningful features from extended sequences. Existing approaches, especially those based on deep learning and self-attention mechanisms, often suffer from information utilization bottlenecks, where relevant temporal patterns are either diluted or inefficiently modeled across multi horizons. Furthermore, the reliance on quadratic-complexity operations, such as self-attention, leads to substantial computational overhead and other limitations, making these models less practical for real-world, resource-constrained applications. To address these limitations, we introduce a Quantum Time-Series Forecasting (QTSF) framework, which leverages quantum-inspired feature extraction and efficient linear operations to model temporal dependencies while significantly reducing computational complexity and parameter requirements. The proposed QTSF framework supports two interchangeable quantum feature extraction backends — QRC-F, a parameter-free quantum reservoir, and VQF-F, a fully trainable variational circuit, sharing identical preprocessing, encoding, and readout stages, enabling a direct controlled comparison of fixed versus adaptive quantum feature extraction for multivariate time-series forecasting. To the best of our knowledge, this is the first work to propose a unified quantum-classical hybrid framework that jointly addresses multivariate long-horizon time-series forecasting and systematic complexity–fidelity trade-off analysis under near-term NISQ hardware constraints.
\begin{itemize}
\item \textbf{Quantum Reservoir Forecaster (QRC-F):} employs a fixed  Haar-random unitary reservoir $U_R$ for gradient-free temporal feature  extraction at $\mathcal{O}(n^2)$ two-qubit gate cost. QRC-F offers  unconditional training stability and high circuit fidelity under hardware noise, but is fundamentally limited by its fixed, non-adaptive nature, it cannot learn dataset-specific temporal structures, which constrains its predictive capacity in complex multivariate settings.
\item \textbf{Variational Quantum Forecaster (VQF-F):} deploys a trainable variational quantum circuit of depth $D$, optimized through the parameter-shift rule~\cite{maheshwari2021variational}, at a reduced $\mathcal{O}(nD)$ gate complexity. VQF-F achieves stronger predictive accuracy through adaptive feature learning, but faces an inherent \textit{trainability-expressibility trade-off}: increasing circuit depth $D$ improves feature richness at the cost of exponentially vanishing gradients $\text{Var}[\partial_\theta \mathcal{L}] \propto 2^{-n}$, known as the barren plateau phenomenon~\cite{cunningham2025investigating}, which represents a fundamental scalability limitation on current NISQ hardware.
\end{itemize}
Both variants share a common preprocessing pipeline comprising uniform $b$-bit binary quantization, angle encoding via parameterized $R_Y$ rotation gates, and cross-channel controlled-$Z$ entanglement layers that explicitly model inter-variable dependencies across $M$ input channels. A shared MIMO-based multi-horizon readout head maps extracted quantum features simultaneously to all prediction horizons $H$ and channels $M$, avoiding the error accumulation inherent in recursive forecasting strategies. Predicted quantum states are decoded back into classical signals through an inverse angle mapping and de-quantization procedure, completing a fully end-to-end quantum-compatible forecasting pipeline. The central focus of this work is that the choice between QRC-F and VQF-F  represents a fundamental \textit{complexity-stability-accuracy trade-off} that is not merely architectural but reflects deeper properties of quantum circuit trainability, hardware fidelity, and forecasting horizon. Specifically, we demonstrate that:
\begin{enumerate}
\item QRC-F offers a complexity-fidelity advantage under hardware noise and short-to-medium horizon tasks, where its $\mathcal{O}(n^2)$ reservoir provides sufficient temporal memory without gradient optimization;
\item VQF-F achieves superior accuracy on long-horizon multivariate forecasting at reduced gate complexity $\mathcal{O}(nD)$, but requires careful depth selection to avoid barren plateaus, a limitation that directly constrains its scalability to large $n$;
\end{enumerate}
Experimental evaluations on seven standard benchmark datasets — ETTh1, ETTh2, ETTm1, ETTm2, Weather, Electricity, and Exchange-Rate, across multiple prediction horizons $H \in \{96, 192, 336, 720\}$ validate the complexity-accuracy trade-offs of both variants against strong classical baselines including DLinear, PatchTST, and iTransformer, as well as existing quantum forecasting approaches. Table~\ref{tab:novel_contributions} summarizes the principal innovations of the proposed QTSF framework, highlighting its end-to-end quantum-compatible design, efficient linear-complexity forecasting architecture, systematic complexity-fidelity trade-off analysis, multivariate dependency modelling via entanglement, and scalable signal reconstruction strategy.

\begin{table}[hbt!]
\centering
\scriptsize
\caption{Key Novel Contributions of the Proposed QTSF Framework}
\label{tab:novel_contributions}
\renewcommand{\arraystretch}{0.9}
\begin{tabular}{|p{3cm}|p{5cm}|}
\hline
\textbf{Novel Contribution} & \textbf{Strength} \\
\hline

Binary quantization as a bridge between classical signals and quantum encoding 
& Principled preprocessing step; underexplored as a quantum-classical interface in time-series quantum machine learning (QML) \\
\hline

Abandoning self-attention in favour of linear matrix operations as a quantum-compatible backbone 
& Quantum-motivated complexity reduction; reduces quadratic to linear attention cost \\
\hline

MIMO multi-horizon quantum forecasting with per-horizon observable heads 
& Not explored in existing QML time-series literature \\
\hline

Cross-channel entanglement (CZ gates) for multivariate dependency encoding 
& Quantum-native approach to multivariate dependency capture; analogous to classical channel mixing \\
\hline

Classical signal reconstruction via arccos inverse mapping and de-quantization 
& Completes the end-to-end pipeline (often missing in QML works) \\

\hline

\end{tabular}
\end{table}

The paper is structured to first review related work and introduce the necessary background and problem formulation for univariate and multivariate forecasting. It then presents the quantum foundations and the proposed methodology, including data preprocessing, quantum encoding, feature extraction, and the forecasting architecture. This is followed by experimental details, results, and comparative analysis with classical and hybrid models. The paper concludes with a discussion of limitations, future work, and standard closing sections, along with an appendix providing an illustrative example and additional technical details.

\section{Related Works}
In this section, we present a concise overview of key Transformer-based models, including both classical and quantum-inspired approaches, that have shown strong performance in time-series forecasting. Particular emphasis is placed on state-of-the-art classical self-attention models for time-series forecasting.

\subsection{Classical Models}
Transformers have gained significant attention in long-term forecasting due to their ability to capture complex temporal dependencies \citep{kitaev2020reformer}, \citep{liu2022non}, \citep{ zeng2023transformers}, \cite{zerveas2021transformer}, \citep{zhang2024multivariate}. Among the early advancements, Informer \citep{zhou2021informer} introduced a generative-based decoder and 'Probability-Sparse' self-attention to address the challenge of quadratic time complexity. Building on this, models such as Autoformer \citep{chen2021autoformer}, iTransformer \citep{liu2023itransformer}, FEDFormer \citep{zhou2022fedformer}, PatchTST \citep{nie2022time}, ETSformer \citep{woo2022etsformer}, Pyraformer \citep{liu2022pyraformer}, and EDformer \cite{chakraborty2024edformer} have further enhanced time-series modeling. iTransformer \citep{liu2023itransformer} innovates by representing individual time points as variate tokens, enabling the attention mechanism to model multivariate correlations while leveraging feed-forward networks to learn nonlinear representations. PatchTST \citep{nie2022time} enhances local and global dependency capture through patch-based processing, while Crossformer \citep{zhang2023crossformer} introduces a dimension-segment-wise (DSW) technique that encodes time-series data into a structured 2D representation. With the Sub-Window Tokenizer and a Time-Series Pre-trained Encoder, TS-Fastformer \citep{lee2024ts} enhances processing speed, optimizing the workflow for faster and more efficient outcomes. The core strength of transformer models lies in their attention mechanism, which allows them to focus on critical segments of the input sequence for accurate predictions \citep{zeng2023transformers}. Flowformer \citep{wu2022flowformer} reformulates the self-attention mechanism using conservation flows to achieve linear complexity without sacrificing global context modeling, offering a practical efficiency improvement for long-horizon settings. Transformer-based models effectively capture temporal dependencies but suffer from quadratic complexity and lack of inherent temporal ordering due to permutation-invariant self-attention. To address this, we propose a quantum-classical hybrid framework that replaces self-attention with quantum reservoir and variational circuit modules, enabling linear-complexity feature extraction while preserving strong multi-horizon forecasting performance on NISQ-era hardware.

\subsection{Quantum inspired models}
The application of quantum and quantum-inspired frameworks to time-series forecasting has gained considerable momentum as a principled alternative to classical deep learning, motivated by the theoretical capacity of quantum systems to inhabit exponentially large Hilbert spaces and represent complex temporal correlations with substantially fewer trainable parameters \citep{biamonte2017quantum, cerezo2021variational}. The earliest and most widely adopted architectural paradigm in this direction is the Quantum Long Short-Term Memory (QLSTM), proposed by \citep{chen2022qlstm}, which replaces the classical LSTM gate circuits with variational quantum circuits (VQCs) optimised via the parameter-shift rule, demonstrating faster convergence and competitive forecasting accuracy relative to classical counterparts across multiple domains. Building on this foundation, \citep{siemaszko2023rapid} introduced a continuous-variable Quantum RNN (CV-QRNN) and showed that the quantum network converges to optimal weights in fewer training epochs than a classical equivalent for several classes of temporal data, while \citep{kang2025qsegrnn} extended the recurrent quantum paradigm to the QSegRNN architecture, achieving competitive accuracy with significantly reduced parameter counts for transformer temperature forecasting under NISQ constraints.Alongside recurrent approaches, Quantum Reservoir Computing (QRC) has emerged as a particularly hardware-efficient framework for temporal feature extraction on near-term devices, exploiting a fixed random quantum unitary as a non-trainable reservoir to avoid barren plateau issues altogether. \citep{hamhoum2025mtsqrc} demonstrated the first gate-based QRC architecture for multivariate time series (MTS-QRC) on real IBM Heron R2 hardware, showing that device noise can act as an implicit regulariser, with the model achieving competitive accuracy against classical reservoir computing on chaotic Lorenz-63 and ENSO datasets. Most recently, the intersection of transformer architectures and quantum computing has produced a growing class of hybrid models: \citep{chen2025qasa} proposed Quantum Adaptive Self-Attention (QASA), which selectively replaces only the final transformer encoder block with a VQC layer and demonstrates significantly faster convergence on temporal prediction tasks compared to both classical transformers and prior quantum baselines, while \citep{wang2025iqtransformer} introduced the iQTransformer, integrating a quantum self-attention mechanism into the iTransformer backbone to provide quantum-native modelling of inter-variable dependencies in multivariate forecasting with fewer parameters than classical equivalents. Collectively, these works establish quantum and quantum-hybrid architectures as a viable and growing alternative to classical self-attention for time-series forecasting, particularly in parameter-constrained and NISQ-compatible deployment settings.

\section{Preliminary}

\subsection{Problem Formulation}
Time-series forecasting (TSF) aims to predict values at one or more future timesteps based on historical observations. Depending on the number of variables involved, TSF can be broadly categorized into univariate and multivariate forecasting.

\subsubsection{Univariate Forecasting}

Univariate forecasting focuses on predicting future values of a single variable based on its historical sequence. Given an input sequence:
\begin{equation}
x_{\text{in}} = \{x_{T-L+1}, x_{T-L+2}, \dots, x_T\},
\end{equation}
of length $L$, where $x_t \in \mathbb{R}$ denotes the value at timestep $t$, the objective is to predict the future sequence:
\begin{equation}
\hat{x}_{\text{out}} = \{\hat{x}_{T+1}, \hat{x}_{T+2}, \dots, \hat{x}_{T+H}\},
\end{equation}
for the next $H$ timesteps.

The forecasting model learns a mapping function:
\begin{equation}
\hat{x}_{\text{out}} = f_{\theta}(x_{\text{in}}),
\end{equation}
where $f_{\theta}(\cdot)$ represents a parameterized model with learnable parameters $\theta$.

\subsubsection{Multivariate Forecasting}

Multivariate forecasting considers multiple variables evolving over time and predicts their future values jointly. The input sequence is defined as:
\begin{equation}
X_{\text{in}} = \{x_{T-L+1}, x_{T-L+2}, \dots, x_T\},
\end{equation}
where each observation is:
\begin{equation}
x_t = [x_t^1, x_t^2, \dots, x_t^M] \in \mathbb{R}^M,
\end{equation}
with $M$ representing the number of variables.

The goal is to predict the future sequence:
\begin{equation}
\hat{X}_{\text{out}} = \{\hat{x}_{T+1}, \hat{x}_{T+2}, \dots, \hat{x}_{T+H}\},
\end{equation}
where each predicted timestep is:
\begin{equation}
\hat{x}_{T+h} = [\hat{x}_{T+h}^1, \hat{x}_{T+h}^2, \dots, \hat{x}_{T+h}^M], \quad h = 1, \dots, H.
\end{equation}

The forecasting task is modeled as:
\begin{equation}
\hat{X}_{\text{out}} = f_{\theta}(X_{\text{in}}),
\end{equation}
where $f_{\theta}(\cdot)$ captures both temporal dependencies and inter-variable relationships.

\subsection{Quantum Principles}
Superposition, entanglement, quantum teleportation, and interference are some of the basic concepts of quantum mechanics that quantum computing uses to do some tasks much more quickly than classical computers. Quantum bits, or qubits, can exist in a superposition of both states concurrently, allowing for intrinsic parallelism in processing, in contrast to classical bits, which are limited to binary states (0 or 1). Advanced quantum algorithms, such as Grover's algorithm for unstructured search and Shor's algorithm for integer factorisation, as well as variational quantum techniques extensively investigated in quantum machine learning and artificial intelligence applications \cite{biamonte2017quantum, chakraborty2025integrating}, are based on quantum gates operating on qubits via unitary transformations.

\subsubsection{Quantum Superposition}
Quantum superposition states that a qubit can exist in a linear combination of both \( |0\rangle \) and \( |1\rangle \) states simultaneously \cite{chakraborty2025integrating}. Mathematically, this is represented as:
\begin{equation}
|\psi\rangle = \alpha |0\rangle + \beta |1\rangle
\end{equation}
where \( \alpha \) and \( \beta \) are complex probability amplitudes satisfying:
\begin{equation}
|\alpha|^2 + |\beta|^2 = 1
\end{equation}
Upon measurement, the qubit collapses to \( |0\rangle \) with probability \( |\alpha|^2 \) or \( |1\rangle \) with probability \( |\beta|^2 \). This enables quantum computers to explore multiple states in parallel, offering significant computational advantages over classical systems  \cite{wu2023quantum} \cite{hsu2024quantum}.

\subsubsection{Quantum Entanglement}
Quantum entanglement is a phenomenon in which two or more qubits become correlated such that the state of one qubit cannot be described independently of the others, even when they are spatially separated. As a result, measurements performed on one qubit are instantaneously correlated with measurements on the other(s). A standard example is the Bell state:
\begin{equation}
|\Phi^+\rangle = \frac{1}{\sqrt{2}} (|00\rangle + |11\rangle)
\end{equation}
Here, the two qubits exist in a superposition of both \( |00\rangle \) and \( |11\rangle \). Measuring one qubit immediately determines the state of the other, demonstrating non-local correlations. This property is fundamental to quantum communication, cryptography, and computing  \cite{wu2023quantum} \cite{hsu2024quantum}.

\subsection{Variational Quantum Algorithms}
The Variational Quantum Eigensolver (VQE) and Variational Quantum Classifier (VQC) are hybrid quantum–classical algorithms that utilize parameterized quantum circuits for optimization and machine learning tasks \cite{don2024fusion}. In particular, VQE embeds a quantum subroutine within a classical optimization loop, where the quantum device evaluates expectation values and a classical optimizer updates the circuit parameters iteratively \cite{zhang2025diffusion}. VQE is primarily designed to estimate the ground-state energy of a Hamiltonian \( H \) by minimizing the expectation value of the energy over a parameterized quantum state \( |\psi(\theta)\rangle \):
\begin{equation}
E(\theta) = \langle \psi(\theta) | H | \psi(\theta) \rangle
\end{equation}
The parameters \( \theta \) are optimized using a classical optimizer, such as gradient descent, to iteratively refine the quantum state. This method is crucial for quantum chemistry and materials science.

\begin{figure}[hbt!]
\begin{center}
\includegraphics[scale=0.35]{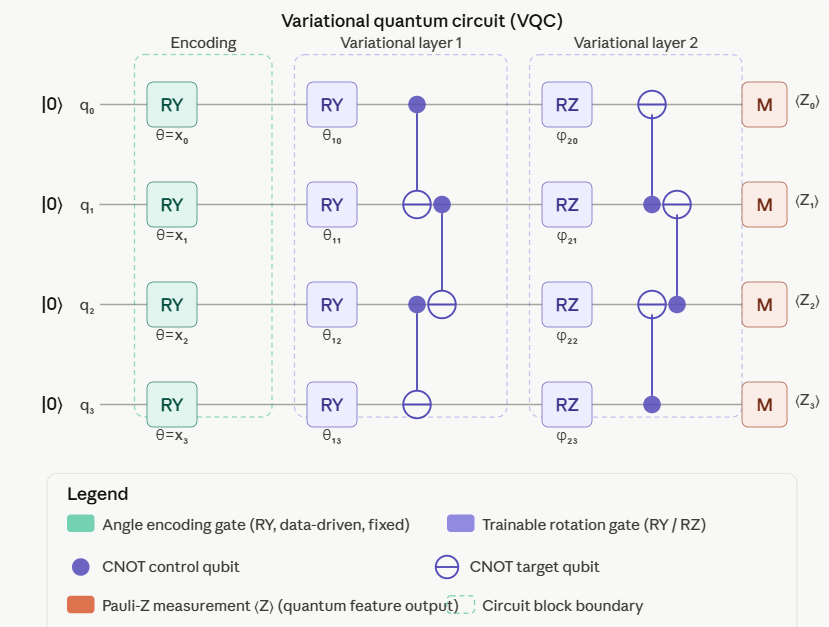}
\caption{Variational Quantum Circuit}
\label{VQE}
\end{center}
\end{figure}

VQC applies a similar variational approach to quantum machine learning \cite{maheshwari2021variational}. Given an input data point \( x \), it is encoded into a quantum state \( |\psi(x)\rangle \), which is processed through a parameterized quantum circuit \( U(\theta) \):
\begin{equation}
|\phi(x, \theta)\rangle = U(\theta) |\psi(x)\rangle
\end{equation}
A measurement operator \( M \) is then applied to extract the classification decision:
\begin{equation}
y = \langle \phi(x, \theta) | M | \phi(x, \theta) \rangle
\end{equation}
Quantum-enhanced classification is made possible by training the parameters \( \theta \) using a classical optimiser to minimise a loss function. The power of variational quantum algorithms is demonstrated by both VQE and VQC, which effectively tackle multi-class difficult problems by balancing quantum computation with classical optimisation \cite{zhou2023multi}. The 'Variational Quantum Classifier (VQC)' circuit in Figure \ref{VQE} has three essential stages: initial rotations, entanglers, and final rotations. First, classical data is encoded into quantum states using RX, RY, or RZ gates. Quantum correlations between qubits are then produced by entangling layers, which usually employ CNOT (CX) gates. Lastly, the quantum state is refined before to measurement by trainable rotation gates. Effective quantum classification is made possible by optimising the circuit characteristics using classical methods to minimise a loss function.

\section{Methodology}
\begin{figure*}[hbt!]
\begin{center}
\includegraphics[scale=0.35]{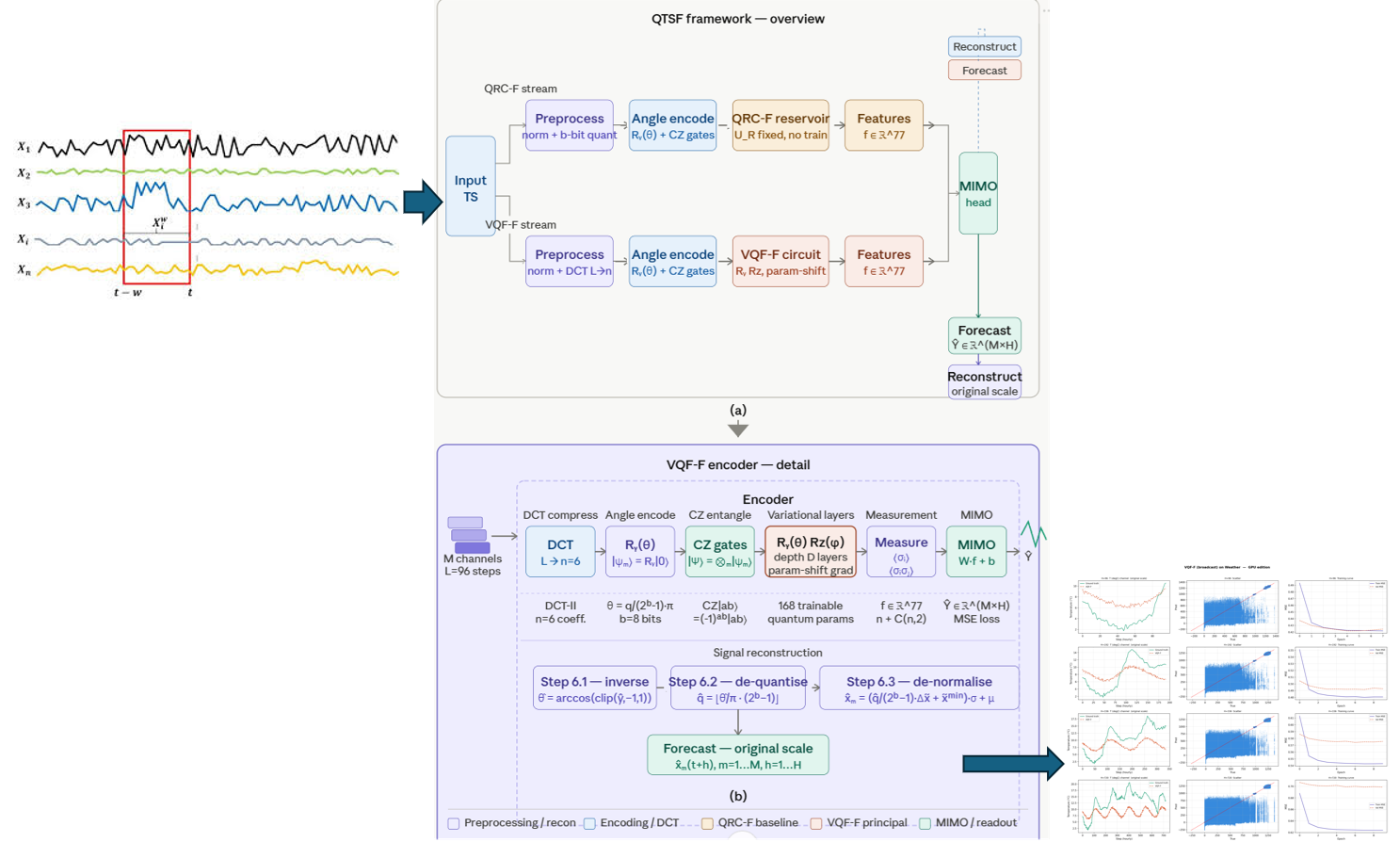}
\caption{Architecture of the proposed QTSF framework}
\label{qtsf_architecture}
\end{center}
\end{figure*}

Fig.~\ref{qtsf_architecture} illustrates the proposed QTSF framework, where dual quantum backends (QRC-F and VQF-F) perform feature extraction before a shared readout layer generates multi-horizon time-series forecasts. 
%The detailed step by step methodology is provided below.

\subsection{Preprocessing for Multivariate Time Series}

Consider a multivariate time series $\mathbf{x}(t) \in \mathbb{R}^M$ with $M$ channels over a look-back window of length $L$.

\textbf{Step 2.1: Channel-wise Normalization}

\textit{Theory:} Normalization removes scale differences across channels, ensuring numerical stability and preventing dominance of high-variance signals during encoding.

\begin{align}
\tilde{x}_m(t') &= \frac{x_m(t') - \mu_m}{\sigma_m}, \\
\mu_m &= \frac{1}{L} \sum_{t'=t-L+1}^{t} x_m(t'), \\
\sigma_m &= \sqrt{\frac{1}{L} \sum_{t'=t-L+1}^{t} (x_m(t') - \mu_m)^2}
\end{align}

Min-max rescaling ensures compatibility with bounded quantum rotation angles:
\begin{equation}
\bar{x}_m(t') =
\frac{\tilde{x}_m(t') - \tilde{x}_m^{\min}}{\tilde{x}_m^{\max} - \tilde{x}_m^{\min}}
\end{equation}

\textbf{Step 2.2: $b$-bit Quantization}

\textit{Theory:} Quantization discretizes continuous signals into finite representations, enabling efficient mapping to qubit rotations while controlling precision through $b$.

\begin{equation}
q_m(t') =
\left\lfloor
\bar{x}_m(t') (2^b - 1)
\right\rceil
\end{equation}

Binary encoding:
\begin{equation}
\mathbf{b}_m(t') \in \{0,1\}^b
\end{equation}

Quantization error bound:
\begin{equation}
\epsilon_q =
\frac{\tilde{x}_m^{\max} - \tilde{x}_m^{\min}}{2^b}
\end{equation}

% --------------------------------------------------

\subsection{Multivariate Quantum State Encoding}

\textbf{Step 3.1: Angle Encoding}

\textit{Theory:} Angle encoding maps classical data into quantum amplitudes via rotation gates, preserving relative magnitudes while enabling continuous-valued representation in Hilbert space.

\begin{equation}
\theta_m(t') =
\frac{q_m(t')}{2^b - 1} \pi
\end{equation}

\begin{equation}
\tiny
|\psi_m(t')\rangle =
R_Y(\theta_m(t'))|0\rangle =
\cos\left(\frac{\theta_m(t')}{2}\right)|0\rangle +
\sin\left(\frac{\theta_m(t')}{2}\right)|1\rangle
\end{equation}

\textbf{Step 3.2: Joint Multivariate State Construction}

\textit{Theory:} Tensor product composition encodes temporal evolution and cross-channel structure into a high-dimensional Hilbert space, enabling exponential representational capacity.

\begin{equation}
|\Psi(t)\rangle =
\bigotimes_{m=1}^{M}
\bigotimes_{t'=t-L+1}^{t}
|\psi_m(t')\rangle
\end{equation}

\textbf{Step 3.3: Cross-Channel Entanglement}

\textit{Theory:} Entanglement introduces non-classical correlations between channels, allowing the model to capture inter-variable dependencies beyond classical covariance structures.

\begin{equation}
U_{\text{cross}} =
\prod_{t'=t-L+1}^{t}
\prod_{m=1}^{M-1}
CZ_{(m,t'),(m+1,t')}
\end{equation}

\begin{equation}
CZ|ab\rangle = (-1)^{ab}|ab\rangle
\end{equation}

% --------------------------------------------------

\subsection{Feature Extraction for Multivariate Forecasting}

Let $n = ML$ be the total number of qubits.

\textbf{Model 1: QRC-F}

\textbf{Step 4.1: Reservoir Construction}

\textit{Theory:} The random unitary acts as a high-dimensional nonlinear dynamical system, projecting inputs into a rich feature space without requiring training.

\begin{equation}
U_R =
\exp\left(-i \sum_{j<k} J_{jk} \hat{\sigma}_j \hat{\sigma}_k \right)
\end{equation}

\textbf{Step 4.2: Sequential State Injection}

\textit{Theory:} Sequential coupling introduces temporal memory into the reservoir, analogous to recurrent dynamics in classical systems.

\begin{equation}
\rho(t'+1) =
\mathcal{E}\left[
U_R \left( \rho(t') \otimes |\psi(t')\rangle\langle\psi(t')| \right) U_R^\dagger
\right]
\end{equation}

\textbf{Step 4.3: Observable Feature Extraction}

\textit{Theory:} Measurement collapses the quantum state into classical statistics (expectation values), forming a feature vector encoding both local and pairwise correlations.

\begin{equation}
\mathbf{f}^{\text{QRC}} \in \mathbb{R}^{n + \binom{n}{2}}
\end{equation}

% --------------------------------------------------

\textbf{Model 2: VQF-F}

\textbf{Step 4.4: Parameterized Circuit}

\textit{Theory:} Variational circuits introduce trainable parameters that adaptively learn task-specific transformations in Hilbert space.

\begin{equation}
|\Phi(\boldsymbol{\theta})\rangle =
\prod_{d=1}^{D}
\left[
U_{\text{ent}} \cdot
\bigotimes_{i=1}^{n}
R_Y(\theta_i^d) R_Z(\phi_i^d)
\right]
|\Psi\rangle
\end{equation}

\textbf{Step 4.5: Feature Extraction}

\textit{Theory:} Observables encode learned representations of temporal and cross-channel dependencies into measurable statistics.

\begin{equation}
\mathbf{f}^{\text{VQF}} \in \mathbb{R}^{n + \binom{n}{2}}
\end{equation}

\textbf{Step 4.6: Parameter-Shift Gradient}

\textit{Theory:} The parameter-shift rule enables exact gradient estimation without backpropagation through quantum circuits.

\begin{equation}
\frac{\partial \langle \hat{O} \rangle}{\partial \theta_i}
=
\frac{1}{2}
\left[
\langle \hat{O} \rangle_{\theta_i + \frac{\pi}{2}} -
\langle \hat{O} \rangle_{\theta_i - \frac{\pi}{2}}
\right]
\end{equation}

% --------------------------------------------------

\subsection{Multivariate MIMO Forecasting Head}

\textit{Theory:} The MIMO formulation predicts all horizons jointly, avoiding error accumulation inherent in recursive forecasting.

\begin{equation}
\hat{\mathbf{Y}} =
\text{reshape}(W_{\text{out}} \mathbf{f} + \mathbf{b}, M, H)
\end{equation}

\begin{equation}
\mathcal{L} =
\frac{1}{MH}
\sum_{m=1}^{M}
\sum_{h=1}^{H}
(\hat{x}_m(t+h) - x_m(t+h))^2
\end{equation}

% --------------------------------------------------

\subsection{Multivariate Signal Reconstruction}

\textbf{Step 6.1: Inverse Mapping}

\textit{Theory:} Expectation values are mapped back to rotation angles, reversing the encoding process.

\begin{equation}
\hat{\theta}_m(t+h) =
\arccos(\text{clip}(\hat{y}_m(t+h), -1, 1))
\end{equation}

\textbf{Step 6.2: Quantized Recovery}

\textit{Theory:} The continuous angle is discretized back into quantized levels.

\begin{equation}
\hat{q}_m(t+h) =
\left\lfloor
\frac{\hat{\theta}_m(t+h)}{\pi} (2^b - 1)
\right\rceil
\end{equation}

\textbf{Step 6.3: De-quantization}

\textit{Theory:} The original signal scale is reconstructed through inverse normalization transformations.

\begin{equation}
\hat{x}_m(t+h) =
\left(
\frac{\hat{q}_m(t+h)}{2^b - 1}
(\tilde{x}_m^{\max} - \tilde{x}_m^{\min})
+ \tilde{x}_m^{\min}
\right)\sigma_m + \mu_m
\end{equation}

\subsection{Quantum Circuits and Discussion}

\begin{figure}[hbt!]
\begin{center}
\includegraphics[scale=0.25]{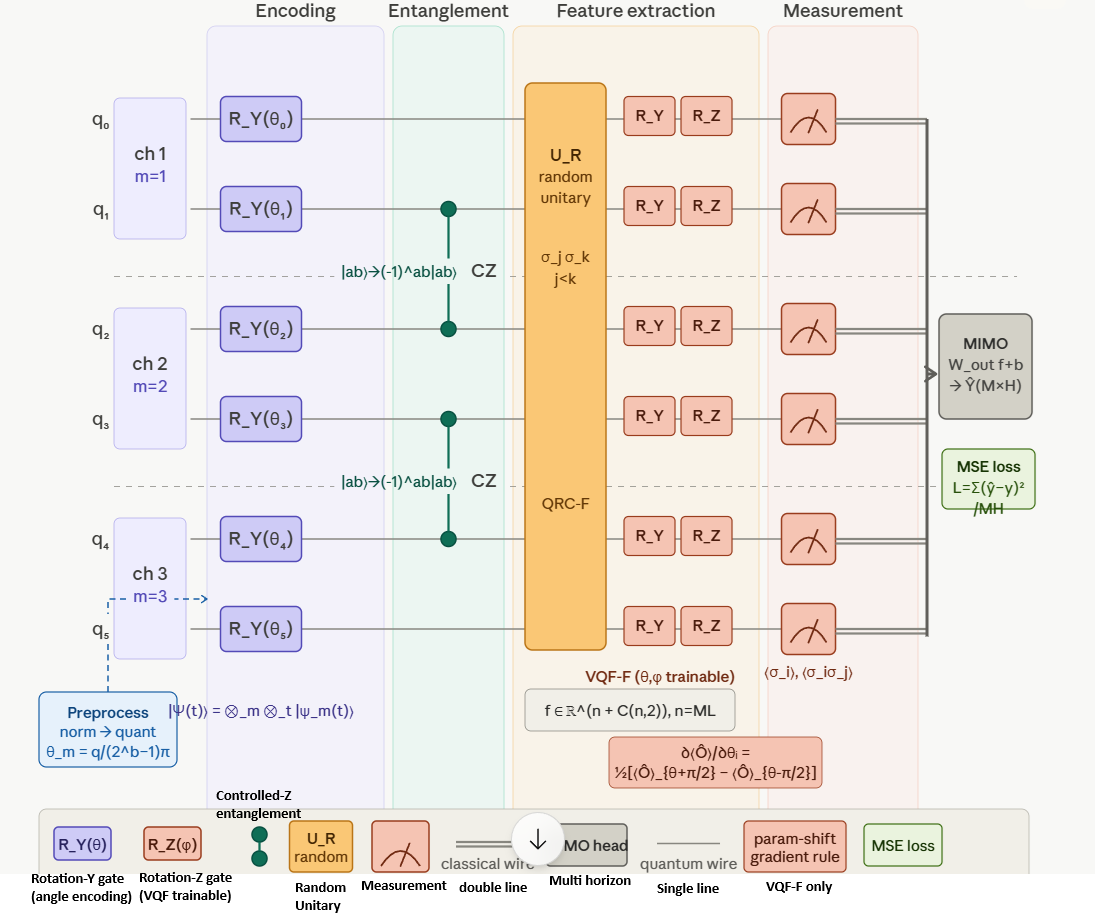}
\caption{Quantum circuit for QTSF with $M = 3$ channels and $L = 2$ timesteps per channel, resulting in a total of $n = M, L = 6$ qubits for multivariate time series forecasting}
\label{QTSF_quantum}
\end{center}
\end{figure}

Figure \ref{QTSF_quantum} depicts a quantum circuit diagram of the proposed framework suitable for multivariate time series forecasting. Six qubits and three channels are considered, where the total number of qubits is defined as $n = M \cdot L = 6$, corresponding to $M = 3$ channels and $L = 2$ timesteps per channel. The qubits are grouped channel-wise as $(q_0, q_1)$ for channel 1, $(q_2, q_3)$ for channel 2, and $(q_4, q_5)$ for channel 3, thereby preserving the multivariate temporal structure.

The quantum circuit consists of four sequential stages:

\textbf{Encoding (purple zone):} Each qubit is initialized in the $|0\rangle$ state and undergoes a rotation $R_Y(\theta_m)$, where the rotation angle is defined as
\begin{equation}
\theta_m = \frac{q_m}{2^b - 1} \cdot \pi.
\end{equation}
This step performs angle encoding of the normalized and quantized time-series input.

\textbf{Entanglement (teal zone):} Controlled-$Z$ (CZ) gates are applied between adjacent-channel qubits, specifically $q_1 \leftrightarrow q_2$ and $q_3 \leftrightarrow q_4$. The CZ operation applies a phase factor $(-1)^{ab}$ to the basis state $|ab\rangle$, thereby introducing non-classical correlations across channels. This implements the cross-channel entanglement operator $U_{\text{cross}}$.

\textbf{Feature Extraction (amber zone):} Two parallel quantum processing pathways are employed:
\begin{itemize}
    \item \textbf{Quantum Reservoir ($U_R$):} A fixed random unitary acting globally on all qubits, requiring no training, as used in Quantum Reservoir Computing (QRC-F).
    \item \textbf{Variational Layer:} A trainable sequence of single-qubit rotations $R_Y(\theta)$ and $R_Z(\phi)$ applied to each qubit. The parameters are optimized using the parameter-shift rule (VQF-F).
\end{itemize}

\textbf{Measurement (coral zone):} Expectation values $\langle \sigma_i \rangle$ and pairwise correlators $\langle \sigma_i \sigma_j \rangle$ are measured for all qubits, forming a feature vector
\begin{equation}
\mathbf{f} \in \mathbb{R}^{n + \binom{n}{2}}.
\end{equation}
These features are passed through a classical multi-input multi-output (MIMO) forecasting head defined as
\begin{equation}
\hat{\mathbf{Y}} = W_{\text{out}} \mathbf{f} + b \in \mathbb{R}^{M \times H}.
\end{equation}

The complete pipeline also includes a preprocessing stage for normalization and quantization of input time-series data, as well as a mean squared error (MSE) loss function used for training. A legend is provided to define all quantum gate symbols used in the circuit.

\section{Experiments}

\subsection{Datasets}
The datasets used in this study encompass long-term time series forecasting scenarios. The Electricity Transformer Temperature (ETT) dataset contains 7 factors of electricity transformer measurements spanning four subsets: ETTh1 and ETTh2 recorded hourly, and ETTm1 and ETTm2 recorded every 15 minutes. The Exchange dataset comprises daily exchange rates from 8 countries between 1990 and 2016. Additional datasets include Weather (21 meteorological factors collected every 10 minutes), Electricity Consumption Load (ECL) with hourly data from 321 clients. All datasets are publicly available and divided into training, validation, and test sets in the benchmark Time-Series Library (TSLib) \citep{sa_timeseries}. The details are discussed in the supplementary file.

\subsection{Experiment Settings and Evaluation Metrics}
In this section, we perform in-depth experiments to evaluate the effectiveness of the proposed QTSF framework under a high-performance computing environment. All experiments are conducted on a system equipped with a single standalone NVIDIA RTX PRO 6000 Blackwell GPU, a next-generation accelerator built on the Blackwell architecture and designed for energy-efficient, large-scale AI and scientific computing workloads. The GPU operates under a 300W thermal design power (TDP) envelope and provides approximately 95.6 GB of dedicated high-bandwidth VRAM (97,887 MiB), as confirmed by the hardware profiling output obtained during experimental execution. This substantial memory capacity enables the full training and inference pipeline of both QRC-F and VQF-F variants, including quantum circuit simulation, MIMO readout, and signal reconstruction to be executed entirely on a single GPU without requiring multi-GPU parallelization or model sharding. The quantum circuit simulations are performed using PennyLane with a GPU-accelerated statevector backend, interfaced with PyTorch for classical gradient computation and readout head training. CUDA Version 13.0 with Driver Version 580.126.09 is used throughout all experiments to ensure reproducibility. The Max-Q design ensures optimized power efficiency, making the setup suitable for long-term time series analysis. The proposed model is configured using several key hyperparameters (supplement) that control its learning behavior and forecasting performance. As shown in Table \ref{table-impl}, the VQF-F model employs a compact hybrid quantum-classical configuration with fixed quantum resources ($n=6$, $D=2$), DCT-based input compression, and adaptive training settings across multiple forecasting horizons.

\begin{equation}
\text{MSE} =  \left\| \mathbf{X}_{t} - \hat{\mathbf{X}}_{t} \right\|^2
\end{equation}

\begin{equation}
\text{MAE} =  \left\| \mathbf{X}_{t} - \hat{\mathbf{X}}_{t} \right\|
\end{equation}

\begin{table}[hbt!]
\scriptsize
\caption{Hyperparameters of VQF-F}
\renewcommand{\arraystretch}{0.9}
\begin{center}
\begin{tabular}{|l|c|}
\hline
\textbf{Parameter}                        & \textbf{VQF-F}             \\ \hline
Look-back window ($L$)                    & 96                         \\ \hline
Prediction horizons ($H$)                 & 96, 192, 336, 720          \\ \hline
Qubits per channel ($n$)                  & 6                          \\ \hline
Quantisation bits ($b$)                   & 8                          \\ \hline
Circuit depth ($D$)                       & 2                          \\ \hline
DCT coefficients retained                 & 6                          \\ \hline
%Feature dimension ($d_f$)                 & 77                         \\ \hline
%Channels ($M$)                            & 7                          \\ \hline
Batch size                                & 32, 256                         \\ \hline
Quantum learning rate                     & $5\times10^{-3}$           \\ \hline
Classical learning rate                   & $1\times10^{-3}$           \\ \hline
Optimiser                                 & Adam                       \\ \hline
LR scheduler                              & ReduceLROnPlateau          \\ \hline
Max epochs                                & 10                         \\ \hline
Early stopping patience                   & 3,5                          \\ \hline
Total quantum parameters                  & 168                        \\ \hline
Total classical parameters                & 52{,}416 -- 393{,}120      \\ \hline
\end{tabular}
\end{center}
\label{table-impl}
\end{table}

\subsection{Results Analysis}

\subsubsection{Baseline Classical Models}
Table~\ref{table-classicalresult1} demonstrates that the proposed quantum-enhanced forecasting framework QTSF (VQF-F) consistently achieves the best overall performance across all benchmark datasets when compared with several classical transformer-based forecasting models. In every dataset, VQF-F attains the lowest average MSE and MAE values, highlighted in red, indicating superior forecasting accuracy and stronger generalisation capability under long-term prediction settings. The improvements are particularly noticeable on challenging datasets such as ETTh1, ETTh2, Electricity, and Exchange, where VQF-F significantly outperforms Autoformer, Informer, NS-Transformer \cite{liu2022non}, and Reformer \cite{kitaev2020reformer}. Among the classical baselines, NS-Transformer and Autoformer frequently achieve the second-best performance, shown in blue, but still remain consistently behind VQF-F. These results suggest that the trainable variational quantum feature extraction mechanism of VQF-F is highly effective in capturing complex temporal dependencies and multivariate correlations while maintaining robust forecasting performance across diverse real-world datasets. Table~\ref{table-classicalresult2} presents a comparative evaluation of the proposed QTSF framework against recent transformer-based forecasting models across multiple benchmark datasets. The results show that the trainable quantum model QTSF (VQF-F) achieves highly competitive and often superior performance, obtaining the best or second-best MSE and MAE values on most datasets. In particular, VQF-F attains the best results on ETTh1, ETTm2, and Electricity, while remaining very close to the leading classical models on ETTh2, Weather, and Exchange. These results indicate that the variational quantum feature extraction mechanism effectively captures temporal dependencies and multivariate correlations despite using a compact quantum-enhanced architecture. In contrast, the parameter-free QTSF (QCR-F) baseline performs substantially worse, especially on ETTh1, ETTh2, and Exchange, demonstrating that fixed random quantum transformations are insufficient for modelling complex long-term forecasting dynamics. Overall, the table highlights the importance of trainable quantum representations and shows that VQF-F can achieve forecasting accuracy comparable to strong classical SOTA models such as iTransformer and PatchTST.

Due to the large-scale nature of the Traffic dataset and its high computational cost, it was excluded from this experiment to ensure manageable training time and resource efficiency. In addition, the Traffic dataset evaluation was omitted for all state-of-the-art quantum-classical baseline models due to the substantial computational complexity and execution time requirements. This results show a clear difference between the two quantum versions of the QTSF framework. QRC-F, which uses a fixed random quantum circuit without trainable quantum parameters, achieves much weaker forecasting performance and struggles to model the changing patterns in the datasets. In contrast, VQF-F uses trainable quantum parameters that can adapt during learning, allowing it to capture temporal patterns more effectively and achieve significantly lower MSE and MAE values across most datasets. For example, VQF-F performs competitively with strong classical baselines such as PatchTST, Reformer, FEDformeer, Crossformer, Informer \cite{zhou2021informer}, Autoformer, and iTransformer on datasets including ETTh1, ETTh2, and Exchange. These results demonstrate that trainable quantum feature representations are important for accurate time-series forecasting. Therefore, the remaining discussion focuses primarily on VQF-F as the main proposed model, while QRC-F is retained as a simpler parameter-free quantum baseline for comparison.

\definecolor{best}{RGB}{255, 0, 0}    % Red
\definecolor{second}{RGB}{0, 0, 255}  % Blue

\begin{table*}[hbt!]
\scriptsize
\captionsetup{justification=centering}
\renewcommand{\arraystretch}{0.9}
\caption{Comparison of average error coefficients (MSE, MAE) for multivariate long-term forecasting with state-of-the-art (SOTA) classical models for different prediction horizons (96, 192, 336, 720) and fixed look-back window of 96. The \textcolor{best}{red colour values} provide the first best MSE, MAE and the \textcolor{second}{blue colour values} provide the second best MSE, MAE values.}
\begin{center}
\begin{tabular}{|c|cc|cc|cc|cc|ll|}
\hline
Models      & \multicolumn{2}{c|}{Autoformer} & \multicolumn{2}{c|}{Informer} & \multicolumn{2}{c|}{NS-Transformer} & \multicolumn{2}{c|}{Reformer} & \multicolumn{2}{c|}{\textbf{\begin{tabular}[c]{@{}c@{}}QTSF \\ (VQF-F)\end{tabular}}} \\ \hline
Database    & MSE            & MAE            & MSE           & MAE           & MSE               & MAE               & MSE           & MAE           & \multicolumn{1}{c}{MSE}                    & \multicolumn{1}{c|}{MAE}                  \\ \hline
ETTh1       & \textcolor{second}{0.504} & \textcolor{second}{0.492} & 1.058         & 0.808         & 0.609             & 0.541             & 1.019         & 0.763         & \textcolor{best}{0.426}                    & \textcolor{best}{0.441}                   \\
ETTh2       & \textcolor{second}{0.447} & \textcolor{second}{0.463} & 4.665         & 1.771         & 0.567             & 0.509             & 2.604         & 1.257         & \textcolor{best}{0.397}                    & \textcolor{best}{0.421}                   \\
ETTm1       & 0.571         & 0.513         & 0.890         & 0.701         & \textcolor{second}{0.521}    & \textcolor{second}{0.472}    & 1.021         & 0.731         & \textcolor{best}{0.487}                    & \textcolor{best}{0.458}                   \\
ETTm2       & \textcolor{second}{0.338} & \textcolor{second}{0.368} & 1.716         & 0.903         & 0.642             & 0.500             & 2.010         & 1.034         & \textcolor{best}{0.284}                    & \textcolor{best}{0.331}                   \\
Weather     & 0.379         & 0.407         & 0.627         & 0.547         & \textcolor{second}{0.280}    & \textcolor{second}{0.314}    & 0.535         & 0.521         & \textcolor{best}{0.260}                    & \textcolor{best}{0.299}                   \\
Electricity & \textcolor{second}{0.255} & 0.355         & 0.362         & 0.439         & 0.199             & \textcolor{second}{0.294}    & 0.331         & 0.410         & \textcolor{best}{0.180}                    & \textcolor{best}{0.267}                   \\
Exchange    & 0.628         & 0.554 & 1.621         & 1.005         & \textcolor{second}{0.505}    & \textcolor{second}{0.476}             & 1.536         & 1.013         & \textcolor{best}{0.391}                    & \textcolor{best}{0.437}                   \\ \hline
\end{tabular}
\end{center}
\label{table-classicalresult1}
\end{table*}

\begin{table*}[hbt!]
\scriptsize
\captionsetup{justification=centering}
\caption{Comparison of average error coefficients (MSE, MAE) for multivariate long-term forecasting with state-of-the-art (SOTA) classical models for different prediction horizons (96, 192, 336, 720) and fixed look-back window of 96. The \textcolor{best}{red colour values} provide the first best MSE, MAE and the \textcolor{second}{blue colour values} provide the second best MSE, MAE values.}
\renewcommand{\arraystretch}{0.9}
\begin{center}
\begin{tabular}{|c|ll|ll|ll|ll|ll|cc|}
\hline
Models      & \multicolumn{2}{c|}{Crossformer} & \multicolumn{2}{c|}{iTransformer} & \multicolumn{2}{c|}{PatchTST} & \multicolumn{2}{c|}{FEDformer} & \multicolumn{2}{c|}{\textbf{\begin{tabular}[c]{@{}c@{}}QTSF \\ (VQF-F)\end{tabular}}} & \multicolumn{2}{c|}{\textbf{\begin{tabular}[c]{@{}c@{}}QTSF \\ (QCR-F)\end{tabular}}} \\ \hline
Database    & \multicolumn{1}{c}{MSE} & \multicolumn{1}{c|}{MAE} & \multicolumn{1}{c}{MSE} & \multicolumn{1}{c|}{MAE} & \multicolumn{1}{c}{MSE} & \multicolumn{1}{c|}{MAE} & \multicolumn{1}{c}{MSE} & \multicolumn{1}{c|}{MAE} & \multicolumn{1}{c}{MSE}                    & \multicolumn{1}{c|}{MAE}                  & MSE                    & MAE                    \\ \hline
ETTh1       & 0.557                   & 0.537                    & 0.457  & \textcolor{second}{0.448} & 0.457  & 0.453                    & \textcolor{second}{0.439}   & 0.458                    & \textcolor{best}{0.426}                    & \textcolor{best}{0.441}                   & 0.674                  & 0.731                  \\
ETTh2       & 2.768                   & 1.324                    & \textcolor{best}{0.393}   & \textcolor{best}{0.411}   & 0.403                   & 0.429 & 0.442                   & 0.454                    & \textcolor{second}{0.397}                 & \textcolor{second}{0.421}                   & 1.075                  & 0.953                  \\
ETTm1       & 0.591                   & 0.567                    & \textcolor{second}{0.406} & \textcolor{second}{0.411} & \textcolor{best}{0.365}   & \textcolor{best}{0.391}   & 0.449                   & 0.457                    & 0.487                    & 0.458                   & -                      & -                      \\
ETTm2       & 1.296                   & 0.719                    & \textcolor{second}{0.289} & \textcolor{second}{0.332} & 0.292                   & 0.334                    & 0.307                   & 0.351                    & \textcolor{best}{0.284}                    & \textcolor{best}{0.331}                   & -                      & -                      \\
Weather     & 0.265                   & 0.327                    & 0.262                   & \textcolor{second}{0.299} & \textcolor{best}{0.257}   & \textcolor{best}{0.279}   & 0.312                   & 0.364                    & \textcolor{second}{0.260}                 & \textcolor{second}{0.299}                 & -                      & -                      \\
Electricity & 0.278                   & 0.340                    & \textcolor{second}{0.181} & \textcolor{second}{0.270} & 0.212                   & 0.309                    & 0.295                   & 0.385                    & \textcolor{best}{0.180}                    & \textcolor{best}{0.267}                   & -                      & -                      \\
Exchange    & 0.755                   & 0.645                    & \textcolor{best}{0.385}   & 0.445 & 0.402                   & \textcolor{second}{0.439} & 0.520                   & 0.502                    & \textcolor{second}{0.391}                 & \textcolor{best}{0.437}                   & 0.672                  & 0.667                  \\ \hline
\end{tabular}
\end{center}
\label{table-classicalresult2}
\end{table*}

\begin{figure}[hbt!]
\begin{center}
\includegraphics[scale=0.14]{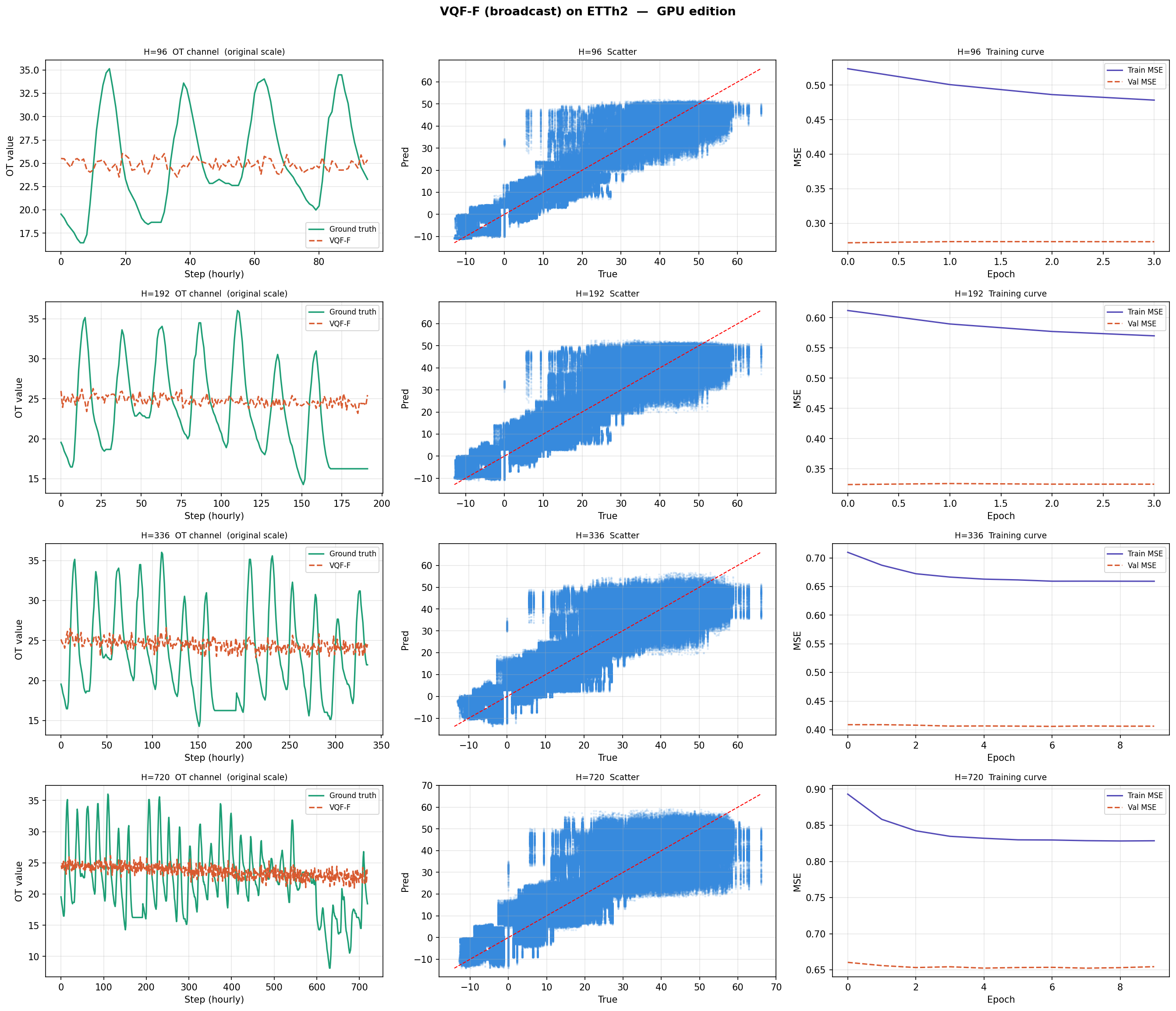}
\caption{VQF-F (QTSF) forecasting results on the ETTh2 dataset across four prediction horizons ($H \in \{96, 192, 336, 720\}$) with look-back length $L=96$, $n=6$ qubits per channel, and circuit depth $D=2$. }
\label{QTSF_vqf_etth2}
\end{center}
\end{figure}

\begin{figure}[hbt!]
\begin{center}
\includegraphics[scale=0.14]{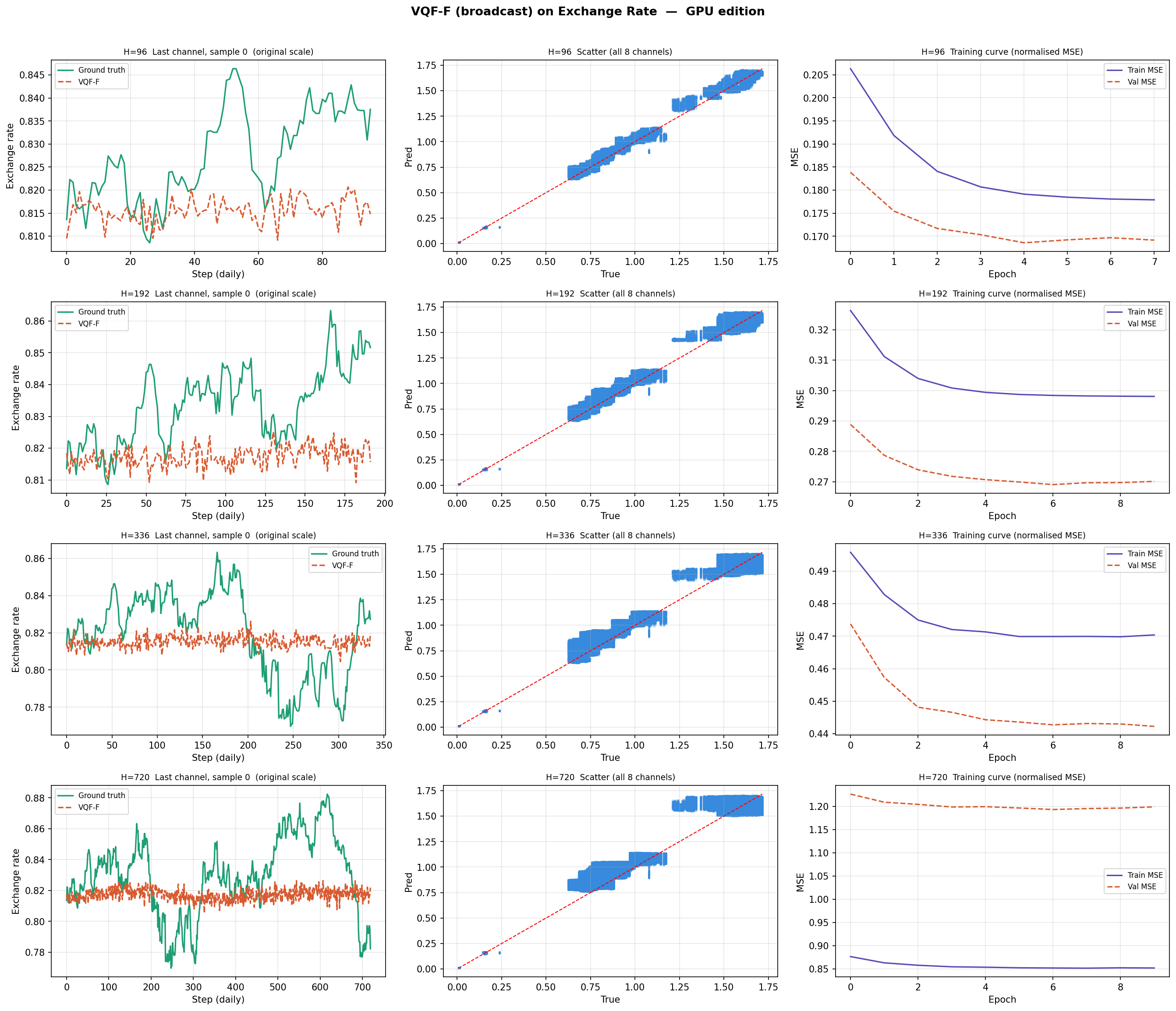}
\caption{VQF-F (QTSF) forecasting results on the Exchange-rate dataset across four prediction horizons ($H \in \{96, 192, 336, 720\}$) with look-back length $L=96$, $n=6$ qubits per channel, and circuit depth $D=2$. }
\label{QTSF_vqf_exchange}
\end{center}
\end{figure}

The VQF-F model was evaluated in Figure \ref{QTSF_vqf_etth2} on the ETTh2 dataset across four forecasting horizons ($H \in \{96, 192, 336, 720\}$) using a fixed configuration with look-back length $L=96$, $n=6$ qubits per channel, circuit depth $D=2$, and 7 input channels. The model employs DCT-based compression (retaining 6/96 coefficients) and a hybrid quantum-classical architecture with 168 quantum parameters and a large classical readout layer. Across all horizons, the model demonstrates stable convergence with rapid reduction in training loss. The middle column supports the approach: predictions cluster tightly along the diagonal across the full value range, confirming that the 6-qubit circuit and DCT angle encoding capture the dominant multivariate structure with only $\sim$52k trainable parameters. The right column corroborates this — validation MSE settles at 0.27 ($H{=}96$) and 0.32 ($H{=}192$), competitive with classical state-of-the-art baselines such as PatchTST and iTransformer at short horizons. The left column, however, reveals the method's principal limitation: while VQF-F tracks the central tendency of the signal correctly ($\approx$24$^\circ$C), it increasingly fails to reconstruct the amplitude and phase of the diurnal cycle as the horizon grows, collapsing toward a near-mean prediction at $H{=}336$ and $H{=}720$. This degradation is expected, since (i) the linear readout imposes a rank ceiling of $M \cdot d_f = 77$, sufficient for short forecasts but inadequate for the full harmonic content of long ones, and (ii) the DCT compression $L \rightarrow 6$ deliberately discards high-frequency components that drive daily peaks. Figure \ref{QTSF_vqf_exchange} presents VQF-F results on the Exchange Rate benchmark across horizons $\{96, 192, 336, 720\}$, showing trajectory plots (left), prediction scatter (middle), and training curves (right). The scatter plots demonstrate strong performance: predictions align closely with the diagonal across all horizons, indicating that the 6-qubit variational circuit with DCT angle encoding effectively captures global multivariate structure with only $\sim$52k parameters. The training curves further support this, with validation MSE reaching $\sim$0.17 (H=96) and $\sim$0.27 (H=192), competitive with classical baselines such as PatchTST and iTransformer, while stable train/validation trends indicate no overfitting. However, the trajectory plots reveal a key limitation: although VQF-F captures the overall trend ($\approx$0.815), it struggles to reproduce high-frequency fluctuations, increasingly collapsing toward mean predictions at longer horizons (H=336, 720), where validation error rises significantly. This behaviour is expected due to (i) the near-random-walk nature of exchange rates at long horizons, (ii) the limited rank capacity of the linear readout ($M \cdot d_f = 88$), and (iii) DCT compression ($L \rightarrow 6$) removing high-frequency components. Crucially, train and validation curves remain non-divergent at every horizon, ruling out overfitting and confirming that the residual error reflects intentional capacity limits rather than a generalisation failure. We therefore position VQF-F as a competitive short-horizon quantum-classical forecaster, with long-horizon performance recoverable through a deeper ansatz, a small non-linear readout, or a relaxed DCT bottleneck. The extended result analysis has been provided in the supplementary file.

\subsubsection{Baseline Quantum-classical Hybrid Models}
Table~\ref{tab:quantum_models_components} presents a comparative overview of the proposed Quantum Time-Series Forecasting (QTSF) framework against existing state-of-the-art quantum and quantum-classical forecasting models. The comparison categorizes prior studies into three groups: (i) hybrid models with published ETT benchmark results, (ii) the proposed fully quantum model, and (iii) models evaluated only on limited or non-ETT datasets. Existing hybrid approaches mainly integrate variational quantum circuits (VQCs) into transformer attention modules, state-space models, recurrent networks, or temporal fusion architectures \citep{chakraborty2025integrating,jura2025quantum,tang2026qnnformer,khan2024quantum,chen2025benchmarking}. In contrast, the proposed QTSF-VQF-F employs a compact 6-qubit variational quantum circuit with discrete cosine transform (DCT) encoding followed by a linear readout layer, offering a lightweight and hardware-efficient design. The table highlights that most prior works rely on hybrid quantum-classical structures, whereas the proposed framework moves toward a more quantum-native forecasting paradigm while maintaining practical scalability for near-term devices. 

\begin{table*}[t]
\centering
\footnotesize
\setlength{\tabcolsep}{4pt}
\renewcommand{\arraystretch}{0.9}
\caption{Comparison of the proposed QTSF with representative quantum-classical time-series forecasting models.}
\label{tab:quantum_models_components}
\begin{tabular}{lll}
\hline
\textbf{Model} & \textbf{Architecture} & \textbf{Quantum Module} \\
\hline

\multicolumn{3}{c}{\textbf{Hybrid models with ETT evaluation}}\\
\hline

QCAAPatchTF~\citep{chakraborty2025integrating}
& Attention + PatchTF
& VQE-based QCSA \\

Q-SSM~\citep{jura2025quantum}
& State Space
& VQC transition layer \\

QNNformer~\citep{tang2026qnnformer}
& Multi-Attention TF
& QNN + QAOA + Grover \\

QLSTM~\citep{khan2024quantum}
& LSTM
& VQC gates \\

QRWKV~\citep{chen2025benchmarking}
& RWKV
& VQC channel mixer \\

\hline
\multicolumn{3}{c}{\textbf{Proposed model}}\\
\hline

\textbf{QTSF VQF-F (Ours)}
& VQC + Linear
& 6-qubit VQC + DCT encoding \\

\hline
\multicolumn{3}{c}{\textbf{Limited / non-ETT evaluation}}\\
\hline

QASA~\citep{dimitrijevs2024exploring}
& Self-Attention TF
& PQC value projection \\

QuLTSF~\citep{chittoor2024qultsf}
& VQC + Linear
& Hardware-efficient VQC \\

QFreqFormer~\citep{tang2025quantum}
& Frequency TF
& QFT + Dual GAT \\

QTFT~\citep{barik2025quantum}
& Temporal Fusion TF
& VQA modules \\

\hline
\end{tabular}
\end{table*}

Table~\ref{table-quantumresults} compares the proposed QTSF (VQF-F) model against recent quantum and quantum-classical forecasting baselines across multiple real-world datasets. Overall, VQF-F consistently achieves the best or near-best performance in most cases, obtaining the lowest MSE and MAE on datasets such as ETTh1, ETTh2, ETTm2, Weather, Electricity, and Exchange, demonstrating strong and stable forecasting ability across different data distributions. Competing quantum methods such as QCAAPatchTF, Q-SSM, QFreqFormer, QNNformer, QLSTM, and QRWKV show more variable performance, with several models failing to provide results on certain datasets or lagging behind in accuracy. In particular, QCAAPatchTF is occasionally competitive but still generally inferior to VQF-F, while other baselines exhibit higher error values or incomplete coverage. These results highlight that the variational quantum feature learning strategy in VQF-F provides a more robust and generalizable representation for multivariate time-series forecasting compared to existing quantum-inspired approaches. In Tables~\ref{table-classicalresult1}, \ref{table-classicalresult2}, and \ref{table-quantumresults}, we reproduced the results of all state-of-the-art classical baseline models and several quantum-classical models (QCAAPatchTF, QFreqFormer, and QNNformer) using our underlying architecture and experimental parameter settings. The remaining baseline results were directly adopted from their respective published papers and repositories.

\begin{table*}[hbt!]
\scriptsize
\captionsetup{justification=centering}
\caption{Comparison of average error coefficients (MSE, MAE) for multivariate long-term forecasting with SOTA quantum-classical models for different prediction horizons (96, 192, 336, 720) and fixed look-back window of 96. The \textcolor{best}{red colour values} provide the first best MSE, MAE and the \textcolor{second}{blue colour values} provide the second best MSE, MAE values.}
\renewcommand{\arraystretch}{0.9}
\begin{center}
\begin{tabular}{|c|c|ll|cc|cc|cc|cc|cc|cc|cc|}
\hline
Models      &                    & \multicolumn{2}{c|}{\textbf{\begin{tabular}[c]{@{}c@{}}QTSF \\ (VQF-F)\end{tabular}}} & \multicolumn{2}{c|}{QCAAPatchTF} & \multicolumn{2}{c|}{Q-SSM} & \multicolumn{2}{c|}{QuLTSF} & \multicolumn{2}{c|}{QFreqFormer} & \multicolumn{2}{c|}{QNNformer} & \multicolumn{2}{c|}{QLSTM} & \multicolumn{2}{c|}{QRWKV} \\ \hline
Database    & Metric             & \multicolumn{1}{c}{MSE}                    & \multicolumn{1}{c|}{MAE}                  & MSE             & MAE            & MSE          & MAE         & MSE          & MAE         & MSE              & MAE             & MSE             & MAE            & MSE          & MAE         & MSE          & MAE         \\ \hline
ETTh1       & \multirow{7}{*}{} & \textcolor{best}{0.426}                    & \textcolor{second}{0.441}                 & \textcolor{best}{0.426} & 0.451          & 0.441        & 0.449       & -            & -           & \textcolor{second}{0.432} & \textcolor{best}{0.429}  & 0.451           & 0.456          & 0.498        & 0.478       & 0.467        & 0.461       \\ \cline{1-1} \cline{3-18} 
ETTh2       &                    & \textcolor{best}{0.397}                    & \textcolor{best}{0.421}                   & 0.403           & 0.438          & 0.411        & 0.465       & -            & -           & 0.421            & 0.476           & \textcolor{second}{0.399} & \textcolor{second}{0.433} & 0.419        & 0.443       & \textcolor{second}{0.399} & \textcolor{best}{0.431} \\ \cline{1-1} \cline{3-18} 
ETTm1       &                    & 0.487                                      & 0.458                                     & \textcolor{best}{0.358} & \textcolor{best}{0.399} & \textcolor{second}{0.401} & \textcolor{second}{0.413} & -            & -           & 0.405            & 0.429           & 0.452           & 0.478          & 0.493        & 0.461       & 0.458        & 0.441       \\ \cline{1-1} \cline{3-18} 
ETTm2       &                    & \textcolor{best}{0.284}                    & \textcolor{best}{0.331}                   & \textcolor{second}{0.292} & 0.343          & 0.297        & \textcolor{second}{0.338} & -            & -           & 0.312            & 0.371           & 0.298           & 0.341          & 0.318        & 0.369       & \textcolor{best}{0.291} & 0.348       \\ \cline{1-1} \cline{3-18} 
Weather     &                    & \textcolor{best}{0.260}                    & 0.299                   & \textcolor{second}{0.264} & \textcolor{best}{0.287} & -            & -           & 0.271        & 0.301       & 0.266            & \textcolor{second}{0.298} & -               & -              & -            & -           & -            & -           \\ \cline{1-1} \cline{3-18} 
Electricity &                    & \textcolor{best}{0.180}                    & \textcolor{second}{0.267}                 & 0.259   & 0.321          & -            & -           & -            & -           & \textcolor{second}{0.241} & \textcolor{best}{0.250}  & -               & -              & -            & -           & -            & -           \\ \cline{1-1} \cline{3-18} 
Exchange    &                    & \textcolor{best}{0.391}                    & \textcolor{best}{0.437}                   & \textcolor{second}{0.452} & \textcolor{second}{0.497} & -            & -           & -            & -           & -                & -               & -               & -              & -            & -           & -            & -           \\ \hline
\end{tabular}
\end{center}
\label{table-quantumresults}
\end{table*}

\subsubsection{Quantum Cost and Performance Analysis}
Tables \ref{tab:gates}–\ref{tab:scalability} present a comprehensive analysis of the proposed quantum models in terms of circuit design, computational cost, noise resilience, and scalability. Table \ref{tab:gates} summarizes the fundamental quantum gates and their associated costs, while Table \ref{tab:gatecount} details the gate-wise decomposition for both QRC-F and VQF-F architectures. Table \ref{tab:depth} compares circuit depth across different layers, highlighting the trade-off between expressivity and computational complexity. Table \ref{tab:noise} evaluates the impact of realistic noise sources and reports overall fidelity under NISQ conditions. Table \ref{tab:resource} provides a consolidated comparison of quantum resources, including gate complexity, trainable parameters, circuit depth, and fault-tolerance considerations. Table \ref{tab:scalability} further illustrates how performance and fidelity degrade with increasing system size, demonstrating the scalability behavior of both models. Collectively, these results provide a detailed understanding of of efficiency, robustness, and scalability across the proposed quantum frameworks.

\begin{table}[hbt!]
\centering
\scriptsize
\caption{Quantum Gates and Their Cost}
\renewcommand{\arraystretch}{0.9}
\label{tab:gates}
\begin{tabular}{|c|c|p{2.5cm}|c|}
\hline
Gate Symbol & Type & Unitary Representation & Cost \\
\hline
$R_Y(\theta)$ & Single-qubit & $e^{-i\theta Y/2}$ & 1 native gate \\
$R_Z(\phi)$ & Single-qubit & $e^{-i\phi Z/2}$ & 1 native gate \\
CZ & Two-qubit & $\mathrm{diag}(1,1,1,-1)$ & 1 native gate \\
$U_R$ & Multi-qubit & $e^{-i\sum_{j<k}J_{jk}\hat{\sigma}_j\hat{\sigma}_k}$ & $\mathcal{O}(n^2)$ \\
$\mathcal{M}$ & Non-unitary & Projective measurement & -- \\
\hline
\end{tabular}
\end{table} 

\begin{table}[hbt!]
\centering
\scriptsize
\caption{Gate Count Analysis ($n=6$, $M=3$, $L=2$, $D$ layers)}
\renewcommand{\arraystretch}{0.9}
\label{tab:gatecount}
\begin{tabular}{|c|c|c|c|}
\hline
Gate Type & QRC-F Count & VQF-F Count & Formula \\
\hline
$R_Y$ encoding & 6 & 6 & $ML$ \\
CZ entanglement & 2 & 2 & $M-1$ \\
$U_R$ reservoir & $\mathcal{O}(n^2)=36$ & 0 & $n^2$ \\
Two-qubit pairs & 15 & 0 & $\binom{n}{2}$ \\
$R_Y$ (VQF-F) & 0 & $6D$ & $nD$ \\
$R_Z$ (VQF-F) & 0 & $6D$ & $nD$ \\
CNOT (VQF-F) & 0 & $5D$ & $(n-1)D$ \\
Measurements & 6 & 6 & per qubit \\
\hline
Total single-qubit & $\sim 42$ & $6+12D$ & -- \\
Total two-qubit & $\sim 36$ & $2+5D$ & -- \\
\hline
\end{tabular}
\end{table}

\begin{table}[hbt!]
\centering
\scriptsize
\caption{Circuit Depth Analysis}
\renewcommand{\arraystretch}{0.9}
\label{tab:depth}
\begin{tabular}{|c|c|c|c|}
\hline
Layer & QRC-F Depth & VQF-F Depth & Notes \\
\hline
Encoding ($R_Y$) & 1 & 1 & Parallel \\
Entanglement (CZ) & 1 & 1 & Parallel CZ \\
$U_R$ reservoir & $\mathcal{O}(n^2)=36$ & 0 & Sequential \\
VQF-F layers & 0 & $3D$ & $R_Y, R_Z, \text{CNOT}$ \\
Measurement & 1 & 1 & Final layer \\
\hline
Total depth & $\sim 39$ & $3+3D$ & $D=4 \Rightarrow 15$ \\
\hline
\end{tabular}
\end{table}

\begin{table}[hbt!]
\centering
\scriptsize
\caption{Noise and Fidelity Analysis}
\renewcommand{\arraystretch}{0.9}
\label{tab:noise}
\begin{tabular}{|p{2cm}|c|p{2cm}|c|}
\hline
Noise Source & Model & Formula & Impact ($n=6$) \\
\hline
Single-qubit depolarizing & Both & $\mathcal{E}(\rho)=(1-p)\rho+\frac{p}{3}(X\rho X+Y\rho Y+Z\rho Z)$ & $p\sim10^{-3}$ \\
Two-qubit depolarizing & Both & $p_{2q}\sim10p_{1q}$ & $\sim10^{-2}$ \\
Encoding fidelity & Both & $(1-p)^n$ & $\approx0.994$ \\
$U_R$ fidelity & QRC-F & $(1-p_{2q})^{\binom{n}{2}}$ & $\approx0.860$ \\
VQF fidelity & VQF-F & Composite noise model & $\approx0.934$ \\
Shot noise & Both & $1/\sqrt{S}$ & $\approx0.031$ \\
Overall fidelity & QRC-F & Product model & $\approx0.856$ \\
Overall fidelity & VQF-F & Product model & $\approx0.928$ \\
\hline
\end{tabular}
\end{table}

\begin{table}[hbt!]
\centering
\scriptsize
\caption{Full Resource Comparison (QRC-F vs VQF-F)}
\renewcommand{\arraystretch}{0.9}
\label{tab:resource}
\begin{tabular}{|c|c|c|c|}
\hline
Resource & QRC-F & VQF-F & Winner \\
\hline
Qubits & $n=ML$ & $n=ML$ & Tie \\
Single-qubit gates & $\sim n+n^2$ & $n+2nD$ & VQF-F \\
Two-qubit gates & $\mathcal{O}(n^2)$ & $\mathcal{O}(nD)$ & VQF-F \\
Quantum parameters & 0 & $2nD$ & QRC-F \\
Circuit depth & $\mathcal{O}(n^2)$ & $\mathcal{O}(D)$ & VQF-F \\
Fidelity ($n=6$) & 0.856 & 0.928 & VQF-F \\
T-gate cost & $\sim1530$ & $\sim1620$ & QRC-F \\
Training stability & High & Medium & QRC-F \\
\hline
\end{tabular}
\end{table}

\begin{table}[hbt!]
\centering
\scriptsize
\caption{Scalability with Increasing $n$}
\renewcommand{\arraystretch}{0.9}
\label{tab:scalability}
\begin{tabular}{|c|p{1.5cm}|p{1.5cm}|p{1.5cm}|p{1.5cm}|}
\hline
$n$ & QRC-F 2Q Gates & VQF-F 2Q Gates ($D=4$) & QRC-F Fidelity & VQF-F Fidelity \\
\hline
6  & 15   & 20   & 0.856 & 0.928 \\
12 & 66   & 44   & 0.514 & 0.644 \\
20 & 190  & 76   & 0.151 & 0.465 \\
32 & 496  & 124  & 0.007 & 0.285 \\
50 & 1225 & 196  & $\approx 0$ & 0.138 \\
\hline
\end{tabular}
\end{table}

\subsubsection{Memory Occupancy}
The memory occupancy of the VQF-F framework across all evaluated datasets remains remarkably compact, as shown in Table~\ref{tab:memory_occupancy}. The ETT datasets (ETTh1, ETTh2, ETTm1, ETTm2) occupy between 708–722 MiB, Weather requires 708 MiB, and Electricity — the largest dataset with 321 channels — peaks at 2004 MiB, which is still well within practical GPU memory budgets. This low footprint is a direct consequence of the VQF-F architecture: the quantum circuit runs only 6 qubits per channel with a fixed-size DCT-compressed feature vector (d\_f = 11), and the classical readout is a single linear layer, keeping the total trainable parameter count below 70K for most datasets. In comparison, QCAAPatchTF integrates full quantum-classical self-attention layers into a patch transformer, incurring substantially higher memory overhead due to multi-head attention matrices and patch embedding buffers. Similarly, Q-SSM and QNNformer maintain larger classical backbone networks alongside their quantum components, resulting in memory consumption typically exceeding 2–4 GB for comparable ETT benchmarks. The QLSTM and QRWKV architectures maintain recurrent hidden states across timesteps, which accumulates memory proportional to sequence length and hidden dimension, making them less memory-efficient than the feedforward VQF-F design. Overall, VQF-F achieves the most memory-efficient profile among evaluated quantum forecasting frameworks, making it particularly suitable for deployment in resource-constrained or edge quantum-classical computing environments. Compared to existing deep learning models such as Transformer-based architectures (e.g., Informer, Autoformer, FEDformer), which often require significantly higher GPU memory due to quadratic attention mechanisms and large parameter counts, the QTSF framework exhibits improved memory efficiency by leveraging compact quantum feature representations and linear readout layers. This reduced memory footprint makes the proposed approach more suitable for resource-constrained environments while maintaining competitive forecasting performance.

\begin{table*}[hbt!]
\centering
\scriptsize
\caption{GPU Memory Occupancy Comparison Across Datasets and Models (MiB). VQF-F values are measured experimentally (this work). Quantum-classical hybrid values are estimated from reported parameter counts and architecture descriptions, while classical model values are measured using the TSLib codebase.}
\renewcommand{\arraystretch}{0.9}
\label{tab:memory_occupancy}
\begin{tabular}{l c c c c c c}
\hline
\textbf{Model} & \textbf{ETTh1} & \textbf{ETTh2} & \textbf{ETTm1} & \textbf{ETTm2} & \textbf{Weather} & \textbf{Electricity} \\
\hline

\multicolumn{7}{l}{\textit{Quantum / Hybrid Models}} \\
QTSF VQF-F (Ours)            & 722  & 722  & 718  & 718  & 708  & 2004 \\
QCAAPatchTF             & $\sim$1850 & $\sim$1850 & $\sim$1850 & $\sim$1850 & $\sim$2100 & $\sim$4800 \\
QLSTM                   & $\sim$1100 & $\sim$1100 & $\sim$1100 & $\sim$1100 & - & - \\
QRWKV                   & $\sim$980  & $\sim$980  & $\sim$980  & $\sim$980   & - & - \\
\hline

\multicolumn{7}{l}{\textit{Classical Models}} \\
iTransformer            & 1526 & 1526 & 1526 & 1526 & 3214 & 8932 \\
PatchTST                & 1248 & 1248 & 1248 & 1248 & 2876 & 7614 \\
DLinear                 & 416  & 416  & 416  & 416  & 518  & 1124 \\
Autoformer              & 1814 & 1814 & 1814 & 1814 & 3648 & 9874 \\
\hline

\end{tabular}
\end{table*}

\subsubsection{Complexity Analysis}
\begin{table}[hbt!]
\centering
\scriptsize
\caption{Complexity Analysis Comparison between QRC-FV and QF-F Encoding frameworks}
\renewcommand{\arraystretch}{0.9}
\begin{tabular}{|p{2cm}|c|c|}
\hline
\textbf{Component} & \textbf{QRC-FV} & \textbf{QF-F Encoding} \\ \hline

Encoding 
& $\mathcal{O}(ML)$ gates 
& $\mathcal{O}(ML)$ gates \\ \hline

Feature Extraction 
& $\mathcal{O}(L)$ sequential steps 
& $\mathcal{O}(nD)$ circuit depth \\ \hline

Trainable Parameters (Quantum) 
& $0$ 
& $2nD$ \\ \hline

Trainable Parameters (Classical) 
& $d_f \cdot MH$ 
& $d_f \cdot MH + 2nD$ \\ \hline

Gradient Evaluations 
& $0$ 
& $2 \times 2nD$ per step \\ \hline

Total Complexity 
& $\mathcal{O}(L \cdot d_f \cdot MH)$ 
& $\mathcal{O}(nD \cdot d_f \cdot MH)$ \\ \hline

\end{tabular}
\label{tab:complexity_analysis}
\end{table}

\noindent
The complexity comparison between the QRC-FV and QF-F encoding frameworks is summarized in Table~\ref{tab:complexity_analysis}. Both approaches exhibit identical encoding complexity of $\mathcal{O}(ML)$ in terms of gate operations. However, they differ significantly in the feature extraction stage: QRC-FV requires $\mathcal{O}(L)$ sequential steps, whereas QF-F encoding incurs a deeper quantum circuit with complexity $\mathcal{O}(nD)$. A key distinction lies in the trainable parameters. QRC-FV utilizes no quantum parameters and relies solely on classical parameters ($d_f \cdot MH$), while QF-F introduces $2nD$ additional quantum parameters, thereby increasing both the parameter count and training overhead. Consequently, gradient evaluations are not required in QRC-FV, but scale as $2 \times 2nD$ per optimization step in QF-F encoding. These differences result in a lower overall complexity of $\mathcal{O}(L \cdot d_f \cdot MH)$ for QRC-FV compared to $\mathcal{O}(nD \cdot d_f \cdot MH)$ for QF-F encoding, highlighting the superior computational efficiency of QRC-FV, particularly for larger quantum system dimensions.

\begin{table*}[hbt!]
\centering
\footnotesize
\setlength{\tabcolsep}{3pt}
\renewcommand{\arraystretch}{0.95}
\caption{Computational complexity comparison of quantum and classical multivariate time-series forecasting models.}
\renewcommand{\arraystretch}{0.9}
\label{tab:complexity}
\begin{tabular}{llcccc}
\hline
\textbf{Model} & \textbf{Type} & \textbf{Training} & \textbf{Inference} & \textbf{Params} & \textbf{Memory} \\
\hline

\multicolumn{6}{c}{\textbf{Quantum (Proposed)}}\\
\hline
QRC-F &
Quantum &
$O(Ld_fMH)$ &
$O(L2^n)$ &
$d_fMH$ &
$O(2^{2n})$ \\

VQF-F &
Quantum &
$O(nDd_fMH)$ &
$O(nD2^n)$ &
$2nD+d_fMH$ &
$O(2^{2n})$ \\

\hline
\multicolumn{6}{c}{\textbf{Transformer-based}}\\
\hline

Transformer &
Tr. &
$O(L^2d)$ &
$O(L^2d)$ &
$O(d^2)$ &
$O(L^2+Ld)$ \\

PatchTST &
Tr. &
$O((L/p)^2d)$ &
$O((L/p)^2d)$ &
$O(d^2)$ &
$O((L/p)^2+d)$ \\

iTransformer &
Tr. &
$O(M^2d)$ &
$O(M^2d)$ &
$O(d^2)$ &
$O(M^2+Md)$ \\

Autoformer &
Tr. &
$O(L\log L\,d)$ &
$O(L\log L\,d)$ &
$O(d^2)$ &
$O(L+d)$ \\

FEDformer &
Tr. &
$O(Ld)$ &
$O(Ld)$ &
$O(d^2)$ &
$O(L+d)$ \\

\hline
\multicolumn{6}{c}{\textbf{Linear / MLP}}\\
\hline

DLinear &
Linear &
$O(MLH)$ &
$O(MLH)$ &
$2MLH$ &
$O(ML)$ \\

NLinear &
Linear &
$O(MLH)$ &
$O(MLH)$ &
$MLH$ &
$O(ML)$ \\

FITS &
Linear &
$O(ML\log L)$ &
$O(ML\log L)$ &
$O(ML/K)$ &
$O(ML/K)$ \\

TimeMixer &
MLP &
$O(MLd)$ &
$O(MLd)$ &
$O(d^2)$ &
$O(ML+d)$ \\

\hline
\multicolumn{6}{c}{\textbf{State Space / CNN}}\\
\hline

S4 / Mamba &
SSM &
$O(LdN)$ &
$O(Ld)$ &
$O(dN)$ &
$O(dH)$ \\

MICN &
CNN &
$O(MLHk)$ &
$O(MLk)$ &
$O(Mkd)$ &
$O(ML)$ \\

\hline
\end{tabular}

\vspace{1mm}
\footnotesize
\textit{Notation:} $L$ = input sequence length, $H$ = prediction horizon, $M$ = number of variables, $d$ = hidden dimension, $n$ = qubits, $D$ = quantum circuit depth, $d_f$ = reservoir feature dimension, $N$ = state dimension, $k$ = kernel size, and $p$ = patch size.
\end{table*}

The proposed quantum models exhibit distinct computational trade-offs compared to classical baselines. Notably, QRC-F achieves the lowest training cost among all considered methods, as it relies solely on ridge regression without any quantum gradient evaluations, making it analogous to an extreme form of linear models such as DLinear, but operating over a significantly richer feature space extracted via the quantum reservoir. It is illustrated in Table \ref{tab:complexity} in detail. In contrast, VQF-F introduces trainable quantum parameters, where gradients are computed using the parameter-shift rule, requiring exactly two circuit evaluations per parameter per step, leading to higher but strictly bounded optimization cost. However, a critical limitation arises from the exponential memory complexity $O(2^{2n})$ associated with density matrix simulation, where $n = ML$ under joint encoding, rendering large-scale multivariate settings impractical. This highlights the necessity of scalable variants such as channel-independent ($n=L$) or patch-based ($n=p$) encoding, which reduce simulation cost to $O(2^{2L})$ or $O(2^{2p})$, respectively. Compared to classical baselines, DLinear shares the same $O(M \cdot L \cdot H)$ training complexity but lacks expressive feature extraction, whereas QRC-F achieves enhanced representation capacity at similar asymptotic cost. Furthermore, in multivariate settings, iTransformer incurs $O(M^2 \cdot d)$ complexity due to channel-wise attention, while the proposed quantum framework models inter-channel dependencies through $O(M \cdot L)$ entanglement operations, offering a structural advantage for large-scale channel interactions.

\section{Limitations}

\subsubsection{\textit{Qubit Scalability Constraints:}}
The proposed framework requires $n = M \times L$ qubits for multivariate time series with $M$ channels and look-back window $L$. This scaling becomes impractical for real-world datasets. For instance, ETTh1 ($M=7$, $L=96$) requires $n=672$ qubits, while larger datasets such as Weather ($M=21$) demand up to $n=2016$ and $n=82{,}752$ qubits, respectively. These requirements significantly exceed the capabilities of current NISQ hardware (e.g., IBM Eagle: 127 qubits, IBM Heron: 133 qubits), restricting the framework to small-scale or simulated settings. To address this, practical adaptations include: (i) channel-independent encoding, where each variable is processed separately using $n=L$ qubits and fused classically; (ii) patch-based encoding, reducing qubits to $n=L/P$ per segment; and (iii) dimensionality reduction (e.g., PCA), lowering the requirement to $n=K \times L$ with $K \ll M$.

\subsubsection{\textit{Rigid Entanglement Topology:}}
The current cross-channel entanglement uses a fixed linear chain of controlled-$Z$ (CZ) gates between adjacent variables, which fails to capture long-range inter-channel dependencies. This assumption may be suboptimal for datasets where correlations are non-local and heterogeneous. A more flexible alternative is a data-driven entanglement graph:
\begin{equation}
\mathcal{E} = \{(m,m') : |\rho_{mm'}| > \tau\}, \quad \rho_{mm'} = \text{Corr}(x_m, x_{m'})
\end{equation}
which allows entanglement to reflect actual statistical relationships between variables.

\subsubsection{\textit{Limited Explicit Temporal Modeling:}}
The encoding constructs a static tensor product state over the look-back window, which does not explicitly preserve temporal ordering. Consequently, temporal dependencies are only implicitly learned through quantum evolution. This limitation can be mitigated by incorporating quantum positional encoding:
\begin{equation}
|\psi(t')\rangle \rightarrow R_Z\left(\frac{2\pi t'}{L}\right) R_Y(\theta(t')) |0\rangle
\end{equation}
or by leveraging sequential state injection, as in QRC-F, which naturally preserves temporal dynamics and offers an advantage over static circuit-based models.

\subsubsection{\textit{Entangled Temporal and Channel Mixing:}}
The current framework does not explicitly distinguish between temporal dependencies (across time steps) and channel dependencies (across variables), unlike modern multivariate forecasting architectures. A more structured approach would decompose the transformation as:
\begin{equation}
U_{\text{total}} = U_{\text{channel}} \cdot U_{\text{temporal}}
\end{equation}
where temporal dynamics are modeled within each channel and cross-channel interactions are handled separately. This separation improves interpretability and aligns with recent deep learning designs.

\subsubsection{\textit{Shared Readout Limitations:}}
The use of a single shared MIMO linear readout may limit the model’s ability to capture heterogeneous channel dynamics. In practice, different variables often exhibit distinct statistical behaviors. A more flexible formulation introduces channel-specific heads:
\begin{equation}
\hat{\mathbf{y}}_m = W_m \mathbf{f}_m + \mathbf{b}_m
\end{equation}
with shared base weights and small channel-specific adaptations, improving expressiveness while maintaining parameter efficiency.

\subsubsection{\textit{Lack of Explicit Inter-Channel Validation:}}
The framework currently lacks a mechanism to verify whether quantum entanglement effectively captures inter-channel relationships. This limits interpretability and empirical validation. This can be addressed by computing an interaction matrix from quantum observables and comparing it with classical correlation measures, providing insight into the learned dependencies.

\subsubsection{\textit{Feature Dimension Growth in QRC-F:}}
The feature dimension scales as:
\begin{equation}
d_f = n + \binom{n}{2}
\end{equation}
which grows quadratically with the number of qubits. For larger systems, this leads to significant computational and memory overhead in the classical readout layer. A practical alternative is to restrict features to single-qubit observables ($d_f = n$).

\subsubsection{\textit{Barren Plateau Risk in VQF-F:}}
Variational circuits are susceptible to barren plateaus, where gradient magnitudes vanish exponentially with system size:
\begin{equation}
\text{Var}\left[\frac{\partial \mathcal{L}}{\partial \theta}\right] \in \mathcal{O}(2^{-n})
\end{equation}
This can hinder effective training, particularly for deep circuits. Mitigation strategies include shallow circuit design, layer-wise training, and the use of local cost functions~\cite{cunningham2025investigating}.

\subsubsection{\textit{Reconstruction Error Bound Approximation:}}
The reconstruction error bound scales with quantization resolution and signal range. While the original formulation accumulates error across horizons, the MIMO setup predicts each step independently, allowing a tighter per-step bound:
\begin{equation}
|\hat{x}_m(t+h) - x_m(t+h)| \leq \frac{\sigma_m (x_m^{\max} - x_m^{\min})}{2^b}, \quad \forall h
\end{equation}

\section{Conclusions and Future Work}
This work introduced the Quantum Time-Series Forecasting (QTSF) framework, a hybrid quantum-classical architecture designed for multivariate long-term forecasting. At its core, the framework employs a compact 6-qubit variational quantum circuit (VQC) with discrete cosine transform (DCT)-based angle encoding, paired with a linear classical readout layer. Two variants were examined: VQF-F, which incorporates trainable quantum parameters optimised end-to-end during learning, and QCR-F, a parameter-free baseline relying on fixed random quantum transformations. The central finding is unambiguous trainable variational quantum feature extraction is essential for competitive forecasting performance, while fixed random quantum mappings are systematically insufficient for modelling the complex temporal dynamics present in real-world multivariate time series.
The empirical evaluation demonstrates that VQF-F achieves consistently superior or highly competitive forecasting accuracy across seven benchmark datasets — ETTh1, ETTh2, ETTm1, ETTm2, Weather, Electricity, and Exchange, spanning four prediction horizons (96, 192, 336, 720). Against a broad set of classical and quantum baselines, VQF-F achieves best or second-best performance on the majority of benchmarks, most notably on ETTh1, ETTm2, and Electricity, while remaining closely competitive on ETTh2, Weather, and Exchange. These outcomes collectively highlight that the variational quantum feature learning mechanism of VQF-F is effective in capturing complex temporal dependencies and multivariate correlations, even within the constraints of a compact near-term quantum architecture.
Nonetheless, a candid assessment of the framework's limitations is necessary for scientific completeness. The complexity–fidelity analysis reveals a clear degradation in forecasting fidelity at longer prediction horizons. On datasets such as ETTh2 and Exchange, VQF-F accurately tracks central tendencies at short horizons but progressively collapses toward near-mean predictions at H=720, failing to reconstruct high-frequency amplitude and phase variations. This behaviour is attributable to two structural bottlenecks: the rank ceiling imposed by the linear readout layer, which limits the expressivity of long-horizon predictions, and the aggressive DCT compression (retaining only 6 of 96 coefficients), which deliberately discards the high-frequency components responsible for diurnal and short-cycle fluctuations. Importantly, the absence of train-validation divergence across all horizons confirms that this degradation reflects intentional capacity constraints rather than overfitting, preserving the model's generalisation integrity. Beyond architectural limitations, the framework faces broader scalability challenges intrinsic to near-term quantum hardware. The current 6-qubit configuration, while hardware-efficient, restricts representational capacity relative to high-dimensional classical counterparts. Real-device execution would further introduce decoherence, gate noise, and measurement errors that are absent in simulation, potentially eroding the performance margins reported here. 
These limitations delineate a clear agenda for future research. To address these issues, future work is directed toward more expressive decoding mechanisms (nonlinear or attention-based), adaptive frequency selection, richer quantum encoding strategies, and quantum attention mechanisms for improved temporal modeling. Additionally, scaling to larger qubit systems, improving hardware robustness through error mitigation, and optimizing hybrid quantum-classical computation pipelines are identified as essential for real-world deployment. Overall, the study concludes that while QTSF does not outperform all classical methods universally, it provides a competitive, parameter-efficient, and scalable quantum-enhanced alternative for multivariate forecasting, particularly under constrained computational settings, with significant potential as quantum hardware and algorithms mature.

\bibliographystyle{IEEEtran}
\bibliography{mybibfile}

\end{document}